\documentclass[aps,prb,twocolumn,showpacs]{revtex4}
\usepackage{graphicx}

\begin{document}

\title{Tailoring the Superconductivity and Antiferromagnetic Order in SrFe$_{2-x}$Rh$_x$As$_2$}

\author{Fei Han, Xiyu Zhu, Peng Cheng, Bing Shen and Hai-Hu Wen}\email{hhwen@aphy.iphy.ac.cn }

\affiliation{National Laboratory for Superconductivity, Institute of
Physics and Beijing National Laboratory for Condensed Matter
Physics, Chinese Academy of Sciences, P. O. Box 603, Beijing 100190,
China}

\begin{abstract}
By doping Rh to the Fe sites in SrFe$_2$As$_2$, superconductivity is
induced when the antiferromagnetic (AF) order is suppressed. The
maximum superconducting transition temperature was found at about 22
K with the doping level of x = 0.25. It is found that the
resistivity anomaly associating with the AF order reveals a sharp
drop for the parent phase and the samples with low doping (x =
0.05), while it evolves into an uprising at a higher doping level (x
$\geq$ 0.1). A general phase diagram is depicted which exhibits the
similarity with that of Co doping. Regarding the close maximum
superconducting transition temperatures in doping Co, Rh and Ir, we
argue that the superconductivity is intimately related to the
suppression of the AF order, rather than the electron-phonon
coupling.
\end{abstract} 
\maketitle

Since the material LaFeAsO$_{1-x}$F$_x$\cite{Hosono} with
superconductivity at 26 K was found, the FeAs-based systems have
received tremendous attention. Just in a short period, many
materials with this unique FeAs-base planes were found, and very
soon, the superconducting transition temperature was promoted above
50 K. It was carried out that most of the parent phase of the FeAs
derivative compounds have an antiferromagnetic (AF) order with a
surprisingly small ordered magnetic moment.\cite{DaiPC} It is now
widely accepted that the superconductivity can be induced when this
AF order is suppressed or destroyed by either doping electrons,
holes or applying a pressure.
\cite{wen,Rotter,CWCh,Pressure1,Pressure2}. Among many systems, the
(Sr,Ba)Fe$_2$As$_2$ (FeAs-122) family has received special attention
since it can be fabricated easily by taking the cation doping, for
example doping the Sr or Ba sites by K, Na
etc.,\cite{Canfield,LuoHQ} or substituting the Fe with Co or
Ni.\cite{Sefat,XuZA,BaNiFeAs} In addition, large scale crystals can
be grown in this structure. The early substitution of Fe ions were
all accomplished with the 3d-transition metals nearby Fe, such as
Co, Ni and Mn. Very recently, new superconductors were discovered
with the substitution of Fe ions with Ru, Ir, and Pd\cite{Ru,Ir,Pd}.
As a complete story, it is worth to substitute Fe ions with other
transition metals such as Rh which locates just below Co and above
Ir in the periodic table of elements. In this paper, we report the
successfully fabrication of the new system
SrFe$_{2-x}$Rh$_{x}$As$_{2}$ with a maximum T$_c$ of 21.9 K at about
x = 0.25. The X-ray diffraction pattern (XRD), resistivity, DC
magnetic susceptibility, and upper critical field will be presented
in pursuing a general phase diagram for
SrFe$_{2-x}$Rh$_{x}$As$_{2}$.

The polycrystalline samples SrFe$_{2-x}$Rh$_x$As$_2$ were
synthesized by using a two-step solid state reaction method.
Firstly, SrAs, RhAs and FeAs powders were obtained by the chemical
reaction method with Sr pieces, Rh powders (purity 99.95\%), Fe
powders (purity 99.99\%) and As grains. Then they were mixed
together in the formula SrFe$_{2-x}$Rh$_{x}$As$_{2}$, ground and
pressed into a pellet shape. All the weighing, mixing and pressing
procedures were performed in a glove box with a protective argon
atmosphere (both H$_2$O and O$_2$ are limited below 0.1 ppm). The
pellet was sealed in a silica tube with 0.2 bar of Ar gas and
followed by heat treatment at 900 $^o$C for 50 hours. Then it was
cooled down slowly to room temperature.

 The x-ray diffraction (XRD) measurement was performed
at room temperature using an MXP18A-HF-type diffractometer with
Cu-K$_{\alpha}$ radiation from 10$^\circ$ to 80$^\circ$ with a step
of 0.01$^\circ$. The analysis of x-ray powder diffraction data was
done by using the software of Powder-X\cite{DongC}, and the lattice
constants were derived by having a general fitting (see below). The
DC magnetization measurements were done with a superconducting
quantum interference device (Quantum Design, SQUID, MPMS7). The
zero-field-cooled magnetization was measured by cooling the sample
to 2 K at zero field, then a magnetic field was applied and the data
were collected during the warming up process. The field-cooled
magnetization data was collected in the warming up process after the
sample was cooled down to 2 K at a finite magnetic field. The
resistivity measurements were done with a physical property
measurement system PPMS-9T (Quantum Design) with the four-probe
technique. The current direction was changed for measuring each
point in order to remove the contacting thermal power.

\begin{figure}
\includegraphics[width=8cm]{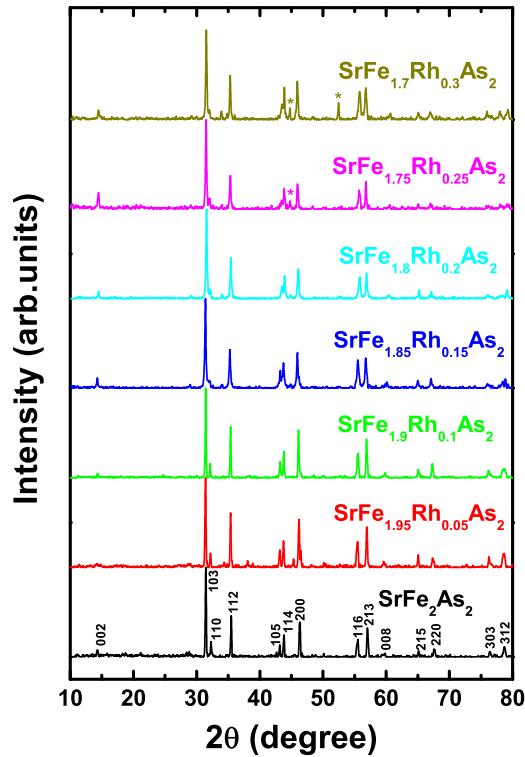}
\caption{(Color online) X-ray diffraction patterns of the samples
SrFe$_{2-x}$Rh$_{x}$As$_{2}$. Almost all main peaks can be indexed
to the tetragonal structure yielding the values of lattice
constants. The asterisks mark the peaks arising from the impurity
phase. } \label{fig1}
\end{figure}

\begin{figure}
\includegraphics[width=8cm]{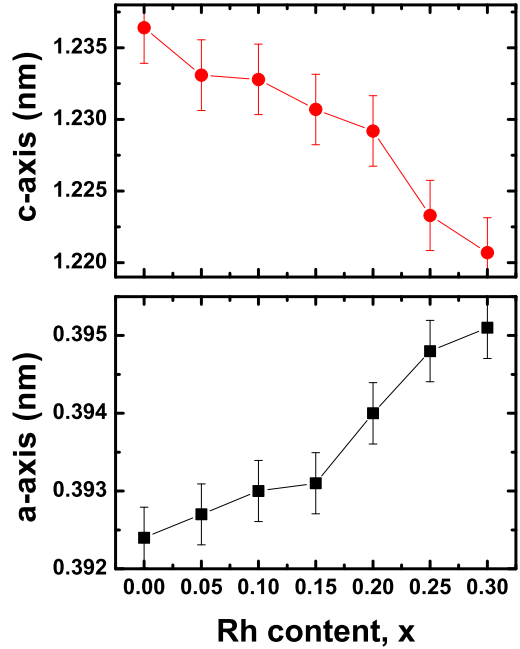}
\caption{(Color online) Doping dependence of the a-axis lattice
constant (top panel) and c-axis lattice constant (bottom panel). It
is clear that the a-axis lattice constant expands, while the c-axis
one shrinks monotonically with Rh substitution. This systematic
evolution clearly indicates that the Rh ions have been successfully
substituted into the Fe-sites. } \label{fig2}
\end{figure}

In Figure 1, we present the x-ray diffraction patterns for
SrFe$_{2-x}$Rh$_{x}$As$_{2}$ with x from 0 to 0.3. All main peaks of
the samples can be indexed to the tetragonal structure very well. In
the sample of x=0.30, there are some tiny peaks arising from the
impurity phase. For all other samples, the impurity phases are
negligible. By fitting the XRD data to the structure calculated with
the software Powder-X,\cite{DongC} we get the lattice constants of
SrFe$_{2-x}$Rh$_{x}$As$_{2}$. In Figure 2, the a-axis and c-axis
lattice parameters for the SrFe$_{2-x}$Rh$_{x}$As$_{2}$ samples were
shown. It is clear that the c-axis lattice constant shrinks, while
the a-axis one expands when more Rh content are doped into the
system. This feature general resembles that in doping Ir, Ru and Pd
into the system.\cite{Ru,Ir,Pd} Normally a smaller a-axis and larger
c-axis lattice constants would mean that the bond angle of As-Fe-As
is smaller. A further refinement of the structural data is underway.

\begin{figure}
\includegraphics[width=8cm]{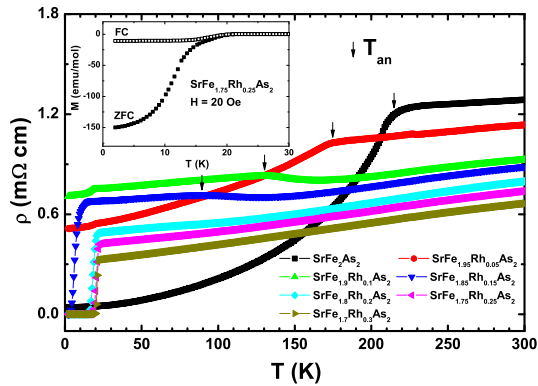}
\caption{(Color online) Temperature dependence of resistivity for
samples SrFe$_{2-x}$Rh$_{x}$As$_{2}$ with x from 0 to 0.3. The inset
shows the temperature dependence of the DC magnetization for the
sample SrFe$_{1.8}$Rh$_{0.2}$As$_{2}$. The resistivity anomaly is
indicated by the arrow for each doping.} \label{fig3}
\end{figure}

In Figure 3, we present the temperature dependence of resistivity
for samples SrFe$_{2-x}$Rh$_{x}$As$_{2}$ and the temperature
dependence of the DC magnetization for the sample
SrFe$_{1.75}$Rh$_{0.25}$As$_{2}$. The parent phase exhibits a sharp
drop of resistivity (resistivity anomaly) at about 215 K. As we can
see, with more Rh doped into the SrFe$_{2-x}$Rh$_{x}$As$_{2}$, the
temperature of this anomaly was suppressed (see, for example the
sample x=0.05). While the anomaly appears still as a sharp drop of
resistivity. When x increases to 0.1 or a higher value,
superconductivity appears, while the anomaly still exists. But now
the resistivity anomaly shows up as uprising. This is slightly
different from the case in Co doping, where a very small amount of
Co doping will convert this sharp drop to an uprising. In the sample
of x = 0.2, the the resistivity anomaly disappeared completely. With
x = 0.25, the maximal T$_c$ with 21.9 K was found. The maximal
transition temperature appears at a higher doping level here
(x=0.25) compared with the case of doping Co (x=0.10-0.16). The
underlying reason is unknown yet. However it is interesting to
mention that in the Ir-doped case, the maximal T$_c$ appears at
about x=0.45. It is yet to be understood whether this is due to the
evolution from doping with 3d (Co), 4d (Rh) and 5d (Ir) transition
metals. The superconducting transition temperatures in this paper
were determined by a standard method, i.e., using the crossing point
of the normal state background and the extrapolation of the
transition part with the most steep slope. A typical example is
given in Fig.5 by the dashed lines for the sample x=0.25 at zero
field. The inset of Figure 3 shows the temperature dependence of the
DC magnetization for the sample SrFe$_{1.75}$Rh$_{0.25}$As$_{2}$.
The measurement was done under a magnetic field of 20 Oe in
zero-field-cooled and field-cooled modes. In an enlarged view near
T$_c$, a clear diamagnetic signal appears below 21$\;$K,
corresponding to the middle transition temperature of the
resistivity data.

\begin{figure}
\includegraphics[width=8cm]{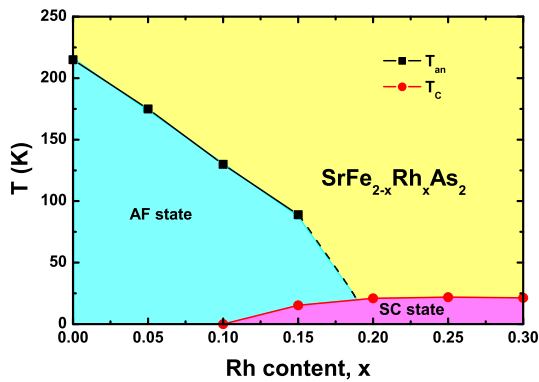}
\caption {(Color online) Phase diagram of
SrFe$_{2-x}$Rh$_{x}$As$_{2}$ within the range of x = 0 to 0.3. The
temperature of resistivity anomaly represents the onset of kink in
resistivity-temperature curve, which corresponds to the
antiferromagnetic order/strctural transitions. The superconductivity
starts to appear at x = 0.1, reaching a maximum T$_c$ of 21.9 K at
about x = 0.25. The dashed line provides a guide to the eyes for the
possible AF order/strctural transitions near the optimal doping
level.} \label{fig4}
\end{figure}

Based on the measurements described above, we can get an electronic
phase diagram for SrFe$_{2-x}$Pd$_{x}$As$_{2}$ within the range of x
= 0 to 0.3, which is shown in Figure 4. Both T$_{an}$ and T$_{c}$
was defined as the temperature at which the anomaly appears in
resistivity and the superconductivity transition, respectively. Just
like other samples in FeAs-122 phase, with increasing Rh-doping, the
temperature of the anomaly is driven down, and the superconducting
state emerges at x = 0.1, reaching a maximum T$_c$ of 21.9 K at x =
0.25. The superconducting state even appears at the doping level of
0.3. However, from the XRD data, we can see that slight impurities
are showing up in this overdoped sample. In addition, from the
diamagnetization measurements, we found that this sample has a much
smaller superconducting volume compared with that of x=0.25. As one
can see, there exists a region in which the antiferromagnetic and
superconductivity coexists in the underdoped side. This general
phase diagram looks very similar to that of
Co-doping\cite{IRFisher}. Since Rh locates just between Co and Ir in
the periodic table of elements, we would conclude that the
superconductivity induced by Rh doping shares the similarity as that
of Co and Ir doping.

\begin{figure}
\includegraphics[width=8cm]{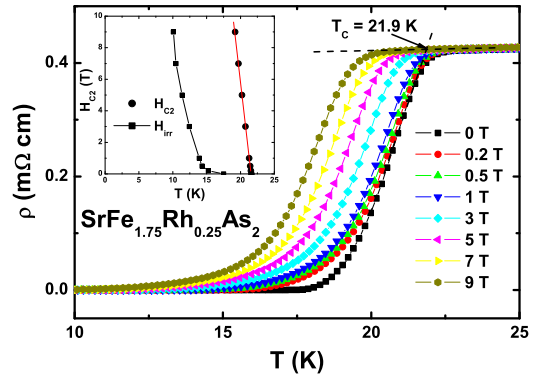}
\caption {(Color online) Temperature dependence of resistivity for
the sample SrFe$_{1.75}$Rh$_{0.25}$As$_{2}$ at different magnetic
fields. The dashed line indicates the extrapolated resistivity in
the normal state. One can see that the superconductivity seems to be
robust against the magnetic field and shifts slowly to the lower
temperatures. The inset gives the upper critical field determined
using the criterion of 90\%$\rho_n$. A slope of -dH$_{c2}$/dT = 3.8
T/K near T$_c$ is found here. The irreversibility line H$_{irr}$
taking with the criterion of 0.1\% $\rho_n$ is also presented in the
inset.} \label{fig5}
\end{figure}

In Figure 5 we present the temperature dependence of resistivity
under different magnetic fields. Just as many other iron pnictide
superconductors, the superconductivity is also very robust against
the magnetic field in the present sample. We used the criterion of
$90\%\rho_n$ to determine the upper critical field and show the data
in the inset of Figure 5. A slope of -dH$_{c2}$/dT = 3.8 T/K can be
obtained here. This is a rather large value which indicates a rather
high upper critical field in this system. By using the
Werthamer-Helfand-Hohenberg (WHH) formula\cite{WHH}
$H_{\mathrm{c}2}(0)=-0.69(dH_{c2}/dT)|_{T_c}T_c$, the value of zero
temperature upper critical field can be estimated. Taking
$T_\mathrm{c}= 21.9\;$K, we can get $H_{\mathrm{c}2}(0) \approx 57.4
T$ roughly. This is a very large upper critical field as in
K-doped\cite{WangZSPRB} and Co-doped samples\cite{Jo}.

The superconductivity mechanism in the FeAs-based superconductors
remains unclear yet. However, our present work and that with the Co
and Ir doping may give some hints on that. First of all, the three
kind of dopants (Co, Rh and Ir) reside in the same column in the
periodic table of elements. The relative atomic mass of these ions
are quite different: 58.9 for Co, 102.9 for Rh and 192.2 for Ir.
Since these atoms are doped into the FeAs-planes, they are certainly
playing important roles in governing the superconductivity. It is
important to note that doping the three different atoms into the
system leads to quite close maximum T$_c$s: 24 K for Co-doping, 22 K
for Rh-doping and 23 K for Ir-doping. In the simple picture
concerning the electron-phonon coupling as the mechanism for the
pairing, the Ir-doped sample should have the lowest T$_c$. We can
even have a brief estimate on T$_c$ based on the electron-phonon
coupling picture. For the Co-doped sample, the maximal T$_c$ appears
at about x=0.16. In this case, we have a average mass
(1.84*55.8+0.16*58.9)/(2Fe) = 56/Fe. Similarly in the Rh doped case,
the maximal T$_c$ appears at about x=0.25, the average mass is
61.7/Fe. For Ir-doping, the maximal T$_c$ appears at about x=0.43,
the average mass is 85.1/Fe. Using the relation of the isotope
effect $M^{\alpha}T_c = constant$ and taking $\alpha=0.5$, we would
have T$_c$ (Co-doping):T$_c$ (Rh-doping):T$_c$ (Ir-doping) = 1: 0.95
:0.81. This is certainly far away from the experimental values.
Although the phonon spectrum as well as the electron band structure
will change with doping Co, Rh and Ir, above argument should be
qualitatively valid. In this sense, the experimental data suggest
that the three elements with very different mass lead to negligible
effect on the superconducting transition temperatures. Actually our
experiment naturally supports the picture that the superconductivity
is established by suppressing the AF order. The related and widely
perceived picture is that the pairing is through the inter-pocket
scattering of electrons via exchanging the AF spin
fluctuations.\cite{Mazin,Kuroki,WangF,WangZD} By doping electrons or
holes into the parent phase, the AF order will be destroyed
gradually. Instead the short range AF order will provide a wide
spectrum of spin fluctuations which may play as the media for the
pairing between electrons. This picture can certainly give a
qualitative explanation to the occurrence of superconductivity in
the cases of doping Co, Rh and Ir.

In summary, superconductivity has been found in
SrFe$_{1-x}$Rh$_{x}$As$_2$ with the maximum T$_c$ = 21.9 K. The
phase diagram obtained is quite similar to that by doping Co or Ir
to the Fe sites. Regarding the close maximal superconducting
transition temperatures in doping Co, Rh and Ir although they have
very different masses, we argue that the superconductivity is
closely related to the suppression of the AF order, rather than the
electron-phonon coupling.

This work is supported by the Natural Science Foundation of China,
the Ministry of Science and Technology of China (973 project:
2006CB601000, 2006CB921802), the Knowledge Innovation Project of
Chinese Academy of Sciences (ITSNEM).

\end{document}